\documentclass[showpacs,twocolumn]{revtex4}%
\input epsf.sty
\usepackage{graphicx}
\usepackage{bm}
\usepackage{amsmath}
\usepackage{amsfonts}
\usepackage{amssymb}%
\begin{document}

\title{Isospin splitting of the nucleon mean field}

\author{W. Zuo$^{1}$, L.G. Cao$^{2}$, B.A. Li$^{3}$, U. Lombardo$^{2,4}$, C.W. Shen$^{5}$}
\affiliation{$^{1}$Institute of Modern Physics, P.O. Box 31,
Lanzhou 730000, China,\\
$^{2}$LNS-INFN, Via Santa Sofia 44, I-95123 Catania, Italy \\
$^{3}$Department of Chemistry and Physics, Arkansas State
University, State University, AR 72467, USA \\
$^{4}$Dipartimento di Fisica dell'Università, Viale Andrea Doria 6, I-95123 Catania,
 Italy \\
$^{5}$China Institute of Atomic Energy, P.O.Box 275(18), Beijing
102413, China\\ }


\begin{abstract}
The isospin splitting of the nucleon mean field is derived from
the Brueckner theory extended to asymmetric nuclear matter. The
Argonne V18 has been adopted as bare interaction in combination
with a microscopic three body force. The isospin splitting of the
effective mass is determined from the Brueckner-Hartree-Fock
self-energy: It is linear acording to the Lane ansatz and such
that $m^*_n > m^*_p$ for neutron-rich matter. The symmetry
potential is also determined  and a comparison is made with the
predictions of the Dirac-Brueckner approach and the
phenomenological interactions. The theoretical predictions are
also compared with the empirical parametrizations of neutron and
proton optical-model potentials based on the experimental
nucleon-nucleus scattering and the phenomenological ones adopted
in transport-model simulations of heavy-ion collisions. The direct
contribution of the rearrangement term due to three-body forces to
the single particle potential and symmetry potential is discussed.
\end{abstract} \maketitle

\section{Introduction}

The study on the role of isospin degree of freedom is in rapid
progress in both nuclear physics and nuclear astrophysics. The
experimental and theoretical research on isospin physics have
received a strong boost because of the construction of more
advanced detectors (such as Magnex\cite{magnex}) and new
radioactive ion beam facilities (project RIA\cite{ria}). A wide
range of rich phenomenologies from nuclei far from the
$\beta$-stability line to strongly asymmetric compound systems
formed in heavy ion collisions (HIC) requires a deep understanding
of the isospin dependence of the in-medium nuclear effective
interaction in a large range of nucleon density and energy. Among
the interesting new physics, a key point is the interplay between
the isospin $T=0$ and $T=1$ components of the effective
interaction as a function of the isospin asymmetry.

Because of the lack of enough empirical information, the most
reliable theoretical tools are the microscopic parameter-free
approaches based on realistic nucleon-nucleon (NN) forces fitting
the experimental phase shifts of the in-vacuum nucleon-nucleon
scattering. One of the most advanced approaches is the
nonrelativistic Brueckner theory. Over the last decade, in fact,
it has been improved in two aspects: its convergence has been
verified at the level of three-body correlations\cite{song:1998}
and the empirical saturation point has been reproduced by
including microscopic three-body forces (TBF)~\cite{zuo:2002}.
Extending the Brueckner calculations to spin and isospin
asymmetric nuclear matter important predictions have been made on
physical quantities, including the symmetry energy, and the spin
and spin-isospin Landau parameters\cite{zuo:2002a,zuo:2003}.

Intimately related to the effective interaction is the nucleon
self-energy, which brings important information on the momentum
dependence of nuclear mean field, effective mass and optical
potential. In the Brueckner-Hartree-Fock (BHF) approximation the
self-energy takes into account not only the interaction of a
nucleon with inert core (pure BHF mean field) but also with core
excitations \cite{mah76,zuo:1999}. The latter is crucial for the
comparison with the experimental nuclear potential and the optical
model potential at low energy.

The isospin splitting of the nucleon self-energy is the main topic
of the present investigation. It has been calculated in wide
ranges of isospin asymmetry, density and energy for the sake of
application in transport simulations of HIC as well as for
structure calculations close to the neutron and proton drip lines.
Despite some results  existing in the literature since long time,
the present work has been stimulated not only by the new
opportunity that we can now study the effects of three-body forces
in the Brueckner theory, but also by the appearance of
relativistic Dirac-Brueckner-Hartree-Fock (DBHF)
calculations\cite{dalen:2004,samma}. Consistent microscopic
predictions could provide strong constraints for phenomenological
approaches, which are still affected by large uncertainties. These
constraints may lead to the need of new parametrizations of the
Skyrme-like interactions for the calculations far from the
beta-stability line.

 The isovector part of the
neutron and proton single particle (s.p.) potentials, i.e., the
symmetry potential, is one of the basic inputs of the transport
models for the collisions of radioactive nuclei. In general, the
shape of the symmetry energy as a function of density is
determined simultaneously by both the momentum and density
dependence of the symmetry potential\cite{li:2002}, therefore the
determination of the momentum dependence of the symmetry potential
is crucial for constraining the high density behavior of symmetry
energy. In the earlier dynamical simulations of HIC, the momentum
dependence of the symmetry potential was seldom taken into
account. Only recently, Das {\it et al.}~\cite{das:2003} has
proposed some simple phenomenological parametrizations for the
momentum dependence of the symmetry potential which have been
adopted in the dynamical simulations of HIC by Li {\it et
al.}~\cite{li:2004a,li:2005,chen04} where it is shown that the
experimental observables such as the neutron-proton differential
flow, the isospin fractionation and the $\pi^-/\pi^+$ ratio etc.,
are quite sensitive to the momentum dependence of the symmetry
potential. Microscopically, the proton and neutron s.p. potentials
and their isospin dependence have been studied in
Ref.~\cite{zuo:1999,zuo:2002a} within the BHF approach. However,
the momentum and density dependence of the symmetry potential was
not discussed in our previous investigations~\cite{zuo:1999,
zuo:2002a}.

In the present paper, we will concentrate on the discussion of the
isospin splitting of the effective mass and the density and
momentum dependence of the symmetry potential, based on the BHF
approximation~\cite{zuo:1999}. Especially we shall compare our
microscopic symmetry potential with the phenomenological ones of
Ref.~\cite{das:2003} and the predictions of the Dirac-Brueckner
method~\cite{dalen:2004,samma}. The present paper is arranged as
follows. In Sec.~II we present a brief introduction of the
Brueckner-Bethe-Goldstone (BBG) theory for G-matrix with a
microscopic TBF, including self-energy and effective mass of
protons and neutrons. The numerical results for the symmetry
potential are reported and discussed in Sec.~III in comparison
with other model predictions. Summary and conclusions are drawn in
Sec.~IV.

\section{Self-consistent BHF approach including a three-body force}
\subsection{BBG equation}

 The Brueckner theory and its extension to include TBFs are
 described elsewhere~\cite{zuo:2002}. Here we simply give a brief
review for completeness. The starting point of the BHF approach is
the reaction $G$-matrix, which satisfies the following isospin
dependent Bethe-Goldstone (BG) equation,
\begin{eqnarray}
G(\rho, \beta, \omega )&=&\\ \nonumber\upsilon_{NN}
&+&\upsilon_{NN} \sum_{k_{1}k_{2}}\frac{ |k_{1}k_{2}\rangle
Q(k_{1},k_{2})\langle k_{1}k_{2}|}{\omega -\epsilon
(k_{1})-\epsilon (k_{2})}G(\rho, \beta, \omega ) \ ,
\end{eqnarray}
where $k_i\equiv(\vec k_i,\sigma_1,\tau_i)$, denotes the single
particle momentum, the $z$-component of spin and isospin,
respectively. $\upsilon_{NN}$ is the realistic nucleon-nucleon
($NN$) interaction, $\omega$ is the starting energy. The asymmetry
parameter is defined as $\beta=(\rho_n-\rho_p)/\rho$, where $\rho,
\rho_n$, and $\rho_p$ denote the total, neutron and proton number
densities, respectively. For the $NN$ interaction, we adopt the
Argonne $V_{18}$ two-body interaction~\cite{wiringa:1995} plus a
microscopic three-body force (TBF)~\cite{grange:1989}. The TBF is
constructed by using the meson-exchange current
approach~\cite{grange:1989} and the most important mesons, i.e.,
$\pi$, $\rho$, $ \sigma $ and $\omega $ have been
considered~\cite{machleidt:1989}. The parameters of the TBF model
have been self-consistently determined so to reproduce the Argonne
$V_{18}$ two-body force  using the one-boson-exchange potential
model~\cite{zuo:2002}. Their values can be found in
Ref.~\cite{zuo:2002}. The TBF contains the contributions from
different intermediate virtual processes such as virtual
nucleon-antinucleon pair excitations, and nucleon resonances ( for
details, see Ref.~\cite{grange:1989}). The TBF effects on the
equation of state (EOS) of nuclear matter and its connection to
the relativistic effects in the DBHF approach have been reported
in Ref.~\cite{zuo:2002}.

In solving the BG equation for the $G$-matrix, the continuous
choice~\cite{mah76} for the auxiliary potential is adopted since
it provides a much faster convergence of the hole-line expansion
than the gap choice~\cite{song:1998}. One advantage of the
continuous choice is that the auxiliary potential has the physical
meaning of the mean field felt by a nucleon during its propagation
between two successive scatterings in nuclear
medium~\cite{sartor:1999}.

The effect of the TBF is included in the self-consistent Brueckner
procedure along the same lines as in Ref.~\cite{grange:1989},
where an equivalent effective two-body interaction $\tilde v$ is
constructed by weighting the third particle in the real TBF by
means of the defect function. So doing, one avoids the difficulty
of solving the full three-body problem. The effective two-body
interaction ${\tilde v}$ can expressed in $r$-space
as\cite{zuo:2002}
\begin{equation}
\begin{array}{lll}
 \langle\vec r_1 \vec r_2| {\tilde v} |
\vec r_1^{\ \prime} \vec r_2^{\ \prime} \rangle =
\\[6mm]\displaystyle
 \frac{1}{4}
{\rm Tr}\sum_n \int {\rm d} {\vec r_3} {\rm d} {\vec r_3^{\
\prime}}\phi^*_n(\vec r_3^{\ \prime}) (1-\eta(r_{13}' ))
(1-\eta(r_{23}')) \\[6mm]
\times \displaystyle W_3(\vec r_1^{\ \prime}\vec r_2^{\ \prime}
\vec r_3^{\ \prime}|\vec r_1 \vec r_2 \vec r_3) \phi_n(r_3)
(1-\eta(r_{13}))(1-\eta(r_{23}))
\end{array}\label{eq:TBF}
\end{equation}
where the trace is taken with respect to the spin and isospin of
the third nucleon. The function $\eta(r)$ is the defect function.
Since the defect function is directly determined by the solution
of the BG equation\cite{grange:1989}, it must be calculated
self-consistently with the $G$ matrix and the s.p. potential
$U(k)$\cite{zuo:2002} at each density and isospin asymmetry. It is
evident from Eq.(\ref{eq:TBF}) that the effective force ${\tilde
v}$ rising from the TBF in nuclear medium is density dependent. A
detailed description and justification of the method can be found
in Ref.~\cite{grange:1989}.

\subsection{Selfenergy}

In the BHF approximation~\cite{zuo:1999} with the TBF, the
selfenergy is made of three terms:
\begin{eqnarray}
\Sigma(k,\varepsilon)=\Sigma_{bhf}+\Sigma_{cpol}+\Sigma_{tbf}
\label{eq:sigma}
\end{eqnarray}
where the first term is the HF  potential with the $G$-matrix as
the effective interaction, the second term is due to the core
polarization~\cite{mah76}, and the third term stems from the
density dependence of the effective force $\tilde v$, i.e., the
TBF rearrangement term $\tilde v$~\cite{Ring}. The first two terms
in the BHF approximation in asymmetric nuclear matter have been
discussed elsewhere~\cite{zuo:1999}. In general, the TBF effect on
the selfenergy within the BHF framework is twofold. First, it
affects the selfenergy via the modification of the $G$-matrix.
This effect has been embodied in the BHF selfenergy, i.e., the
first two terms in Eq.(\ref{eq:sigma}). Second, the density
dependence of ${\tilde v}$ will induce an additional contribution,
i.e., a rearrangement contribution to the selfenergy (third term
in Eq.(3)). The main contribution of the TBF rearrangement can be
formally written in the BHF approximation
\begin{eqnarray}
\Sigma_{tbf} = \frac{1}{2}\sum_{ij}
<ij|\frac{\delta\tilde{v}}{\delta n_k}|ij>_A n_i n_j
\label{eq:sigma-tbf}
\end{eqnarray}
where $n_i$ is the Fermi step function.
 For most results we present below the selfenergy is calculated
in the BHF approximation, and the effect of the TBF is only
restricted to the G-matrix via the BBG equation with the two body
force and the effective TBF. At the end of Sec.~III we will
discuss explicit effect of the rearrangement term of the effective
three-body force on the self-energy and the symmetry potential.

 When calculated on the energy shell the
self-energy gives rise to the single nucleon potential. The
contribution due to the inert core (BHF) for neutrons and protons
is reported in Fig.\ref{unp} as a function of momentum $k$ for
three densities and several isospin asymmetries $\beta\equiv
(\rho_n-\rho_p)/\rho$. The core polarization term mainly
influences the potential at $k < k_F$\cite{mah76} and it will be
neglected in the discussion of this subsection. In order to
explore the isospin effects on the nucleon effective masses
(Sec~II.~C) we split the neutron and proton s.p. potentials into
the contributions from the isospin $T=0$ and $T=1$ channels, i.e.,
\begin{eqnarray}
U_p(k,\beta) = U_p(k,\beta)_{T=0} + U_p(k,\beta)_{T=1}\\
U_n(k,\beta) = U_n(k,\beta)_{T=0} + U_n(k,\beta)_{T=1}
\end{eqnarray}
The isospin behavior of the neutron or proton s.p. potential is a
result of the competition between the $T=0$ and $T=1$ isospin
channels. As discussed in Ref.~\cite{bomba,zuo:1999} the isospin
effect on the EOS of asymmetric nuclear matter is dominated by the
isospin $T=0$ component of the NN interaction. In Fig.~\ref{ut},
we present the contributions from the isospin $T=0$ and $T=1$
channels to the proton and neutron s.p. potentials at $k=0$,
separately, as a function of $\beta$ with respect to their values
in symmetric nuclear matter ($\beta =0$). It is seen that the
variations versus $\beta$ of the $T=0$ components are much larger
than the corresponding $T=1$ components, i.e., $3\sim 5$ times
larger, implying that the $\beta$ dependence of the neutron and
proton s.p. potentials is determined to a large extent by the
$T=0$ component. This is what expected since as increasing the
neutron excess, the $T=0$ interaction between two unlike-nucleons
( vanishing between two like-nucleons) becomes stronger for
protons and weaker for neutrons. The relatively small deviations
of the $T=1$ components of the $U_n$ and $U_p$ from their common
values in symmetric matter is associated to the variations of the
Fermi surfaces in neutron-rich matter. It is also seen from the
figure that the net contribution of the $T=0$ channels stems
almost completely from the $SD$ tensor channel (squares) which is
strongly attractive at relatively low energies while the
contributions from other $T=0$ channels cancel out each other.
This is in agreement with the previous observation for nuclear
symmetry energy~\cite{bomba}. As a consequence, at low momenta,
the proton s.p. potential becomes more attractive and the neutron
one more repulsive going from symmetry nuclear matter ($\beta=0$)
to pure neutron matter ($\beta=1$), as shown in Fig.~\ref{unp}.
According to the experimental data on the phase shifts of
nucleon-nucleon scattering, the attraction of the $SD$ channel
decreases with energy, so that for a given energy the $T=0$
channel contribution to the splitting $U_n-U_p$ becomes equal to
the $T=1$ channel contribution. As a result the isospin splitting
$(U_{T=0}-U_{T=1})/\beta$ vanishes for a given value of momentum
as marked by the crossing point in Fig.~\ref{unp}. This point is
almost independent of both isospin and density. Therefore, the
increase of the proton potential depth vs. asymmetry results in an
increase of the slope as a function of momentum. This behavior
controls the proton and neutron effective mass splitting in
neutron rich matter. From the above discussion it may be concluded
that the isospin behavior of the momentum dependence of the proton
and neutron s.p. potentials which determines the neutron-proton
effective mass splitting, is essentially controlled by the tensor
component of the NN interaction, or say, by the nature of the NN
interaction. The core polarization affects mainly the s.p.
potential in the low momentum range below the Fermi surface. It
gives a repulsive contribution for both the proton and neutron
potentials and weakens the momentum dependence at low
momentum~\cite{mah76,zuo:1999}. The core polarization
contributions to the proton and neutron potential may cancel out
with each other and thus it modifies only slightly the symmetry
potential at low momentum~\cite{zuo:1999}. Therefore inclusion of
the core polarization will not alter our above discussion and
conclusion.

\begin{center}
\begin{figure}[tbh]

\caption{Neutron and proton BHF mean fields at different isospin
asymmetries for three different nucleon densities} \label{unp}
\end{figure}
\end{center}

\begin{center}
\begin{figure}[tbh]

\caption{Isospin splitting the proton and neutron s.p. potentials
at $k=0$: isospin $T=0$ (solid curves) channels, $T=1$ (dashed
curves) channels, $SD$ tensor channel (squares).} \label{ut}
\end{figure}
\end{center}

\subsection{Effective mass}

The nucleon effective mass $m^*$ stems from the non-local nature
of the s.p. potential felt by a nucleon propagating in nuclear
medium. It is determined by the slope of the real part of the
on-shell self-energy $U(k,\epsilon_k)={\rm
Re}\Sigma(k,\epsilon_k)$ in momentum space, i.e.
\begin{eqnarray}
\frac{m^*}{m}=1-\frac{{\rm d} U(k,\epsilon_k)} {{\rm
d}\epsilon_k}.
\end{eqnarray}
 The effective mass itself is momentum dependent,
but usually its value at the Fermi momentum is considered
(hereafter we only discuss the latter). The s.p. energy is
determined by the following momentum-energy relation,
\begin{eqnarray}
\epsilon_k = \frac{k^2}{2m}+ {\rm Re} \Sigma(k,\epsilon_k)
\end{eqnarray}
for a given approximation of the self-energy. It is clear that the
effective mass arises from both the momentum and energy dependence
of the microscopic s.p. potential. It is a different physical
quantity from the Dirac effective mass associated to the medium
modification of the Dirac spinor, which in fact is due to the
self-consistent requirement in the relativistic Dirac-Brueckner
approach as it has been clarified in
Refs.~\cite{jaminon:1989,dalen:2004} and discussed afterwards.

In isospin asymmetric nuclear matter since the momentum-dependence
of the neutron s.p. potential is different from that of the proton
one, the common value of the neutron and proton effective mass is
expected to split into two branches as a function of isospin
asymmetry parameter. The BHF result with the TBF contribution is
displayed in Fig.~\ref{efmas3}.
\begin{figure}[tbh]
\caption{Isospin splitting of the effective mass from a
Brueckner calculation with three-body force.}%
\label{efmas3}%
\end{figure}
Two main features are displayed by the results: the linear
dependence on the asymmetry parameter $\beta$ extended to the
whole asymmetry range, which is a well-known effect\cite{bomba},
and the isospin splitting with $m^*_n > m^*_p$ in neutron-rich
matter. The latter property is related to the increasing
 (decreasing)  slope of the proton (neutron) potential as
discussed in the preceding subsection, but it  has not yet
received a general consensus. In microscopic theories such as the
BHF method and the DBHF approach, the predicted neutron-proton
effective mass splitting is
 such that $m^*_n>m^*_p$ at Fermi surface in neutron-rich
matter. However, within phenomenological models some
parametrizations of the extended Skryme-like effective interaction
may lead to an opposite isospin splitting~\cite{dito,das:2003}. A
complete settlement  of such a controversy will be one of the most
important goals of isospin observables such as the neutron-proton
differential flow and the $\pi^-/\pi^+$ ratio in HIC induced by
radioactive beams.

In order to get a deeper theoretical insight into the
neutron-to-proton splitting of the effective mass, we want to see
how it comes out within the Brueckner many-body theory.
\begin{figure}[tbh]

\caption{Neutron and proton mass vs. asymmetry: separate
contributions from the E-mass and k-mass in units of the bare
nucleon mass. The baryonic density is $\rho=0.17 fm^{-3}$}
\label{mass}
\end{figure}

The off-shell values of the self-energy $\Sigma(k,\epsilon)$
depend separately on the energy and momentum and, as a
consequence, following Mahaux et al. \cite{mah76}, the effective
mass can be split into a product of the two contributions:
\begin{eqnarray}
 \frac{m^{\ast}}{m} = \frac{m_{e}}{m}\frac{m_{k}}{m} \:,
\label{shen-eq4}
\end{eqnarray}
where
 \begin{eqnarray}
 m_{e}(k) = m\left[1-\frac{\partial\Sigma(k,\epsilon)}
 {\partial\epsilon}\right]_{\epsilon=\epsilon_{k}} \:,
\label{shen-eq2} \end{eqnarray}
\begin{eqnarray}
 m_k(k) = m\left[1+\frac{m}{k}\frac{\partial\Sigma(k,\epsilon)}
 {\partial k}\right]_{\epsilon=\epsilon_{k}}^{-1} \:.
\label{shen-eq3}
\end{eqnarray}
 The $k$-mass $m_k$ is related to the
nonlocality of the microscopic mean field in $r$-space. If the
self-energy is energy independent (static limit), then $m_{e}=m$
and the $k$-mass is equal to the effective mass $m^*$. The
$e$-mass describes the nonlocality in time and is related to the
quasi-particle strength by $m_{e}(k)=m/Z(k)$, which gives the
discontinuity of the momentum distribution at the Fermi surface,
and measures the amount of correlations included in the considered
approximation. The two components of the effective mass for
protons and neutrons are plotted in Fig.~\ref{mass}. It is seen
that the isospin splitting is opposite for the two effective
masses $m_k$ and $m_e$. In neutron-rich matter the neutron
$k$-mass becomes larger than the proton one, while the
neutron-proton $e$-mass splitting is reversed. Although the
absolute strengths of the splitting for the two masses are about
the same, the relative splitting of the $k$-mass is much more
pronounced than that of the $e$-mass due to the smaller value of
the $k$-mass. As a consequence, the $k$-mass predominates the
isospin splitting of the total effective mass $m^*$ and leads to
the result $m_n^*>m_p^*$ in neutron-rich nuclear matter. This
result indicates that the effective mass splitting is dominated by
the nonlocality of the microscopic s.p. potentials in spatial
space.

The effective mass so far discussed is also named Schr\"odinger
mass to distinguish from the Dirac mass appearing in the
relativistic mean field theory (RMT) and the relativistic DBHF
approach ~\cite{brown:1987,jaminon:1989,brockmann:1990}. The Dirac
effective mass stems from the self-consistency requirement between
the s.p. wave function and the s.p. spectrum of the Dirac spinor
which is dressed in nuclear medium, and has no any counterpart in
the non-relativistic limit. As discussed in
Ref.~\cite{brown:1987}, the Dirac mass can be traced back to the
effect of the virtual nucleon-antinucleon pair excitations, i.e.,
the in-medium positive-energy spinor is an admixture of the free
negative- and positive-energy spinors.  In
Ref.~\cite{jaminon:1989}, the origin of different effective masses
defined in literatures has been discussed in more details.  It is
shown that in the relativistic framework, an nonrelativistic-type
of effective mass can be introduced based on the corresponding
Schr\"odinger equivalent s.p. potential and it can be compared to
the empirical value extracted from analyses in the framework of
the nonrelativistic optical and shell models. Investigations for
determining the relativistic effective mass in terms of the
momentum dependence of the s.p. energies in the DBHF framework
have been made for symmetric nuclear matter in
Ref.~\cite{brockmann:1990}. Very recently, Van Dalen et
al.~\cite{dalen:2005} have investigated the isospin splitting of
the nonrelativistic-type effective mass obtained from the
Schr\"odinger equivalent s.p. potential within the DBHF approach.
They find that both the dynamical structure (i.e., the
 momentum and density dependence) and the splitting of the
nonrelativistic-type effective mass are satisfactorily consistent
with the predictions of the nonrelativistic BHF approach. The
Dirac mass in neutron-rich matter shows an opposite isospin
splitting of $m_{\rm D,n}^*<m_{\rm D,p}^*$, indicating that the
virtual pair effect on a neutron spinor becomes stronger as the
matter goes to neutron-richer. The nonrelativistic-type of
effective mass in neutron-rich matter derived from the RMT
displays the same behavior of isospin splitting as the Dirac mass,
which is due to the fact that the nonlocal structure of the
self-energy is neglected in the RMT as discussed in
Ref.~\cite{dalen:2005}.

\section{Symmetry Potential and Optical-Model Potential}

The microscopic self-energy in the BHF approach is nonlocal in
space-time coordinates and thus depends on both momentum and
energy. When evaluated on the energy shell, it corresponds to the
empirical optical model potential~\cite{jaminon:1989} :
\begin{eqnarray}
U_{opt}(E)=\Sigma(k(E),E) , \label{optic}
\end{eqnarray}
where $E$ is the incident energy, and $\Sigma(k(E),E)$ the
on-shell self-energy. The momentum $k=k(E)$ is determined by the
mass-shell relation
\begin{eqnarray}
E\,=\,\frac{\hbar^2k(E)^2}{2m} + \Sigma(k(E),E).
\end{eqnarray}

\begin{figure}[tbh]
\caption{BHF isospin symmetry potential vs. momentum $k$ for three
values of density.} \label{f:iso}
\end{figure}

The isovector part of the s.p. potential, which drives the isospin
splitting of the nucleon mean field in asymmetric nuclear matter,
is linearly decreasing with $\beta$ and thus the symmetry
potential can be defined as
\begin{eqnarray}
U_{\mbox{sym}} = \frac{U_n-U_p}{2\beta}
\end{eqnarray}
where $U_n$ and $U_p$ are the s.p. potentials felt by a neutron
and proton in nuclear medium, respectively. In Fig.~\ref{f:iso}
the BHF symmetry potential $U_{\rm sym}$ is displayed as a
function of momentum $k$ for three densities and several isospin
asymmetries. The nucleon-nucleus scattering is not influenced so
much by the core polarization and it can be neglected. It is seen
that the symmetry potential depends strongly upon both density and
momentum. Above the Fermi surface $U_{\rm sym}$ decreases rapidly
as a function of momentum and saturates at high enough momenta. In
the momentum region relevant to the intermediate HIC up to a beam
energy about 300 MeV per nucleon $U_{\rm sym}$ is positive,
implying that its effect is repulsive on neutrons and attractive
on protons. At higher densities the repulsion (attraction) on
neutrons (protons)
 becomes stronger. In Fig.5 it is also shown that $U_{\rm sym}$
is almost independent of the isospin asymmetry $\beta$, which
indicates that the linear dependence of the neutron and proton
s.p. potentials on $\beta$ persists  at any energy and thus it
provides a microscopic support of the empirical assumption of the
Lane potential~\cite{lane}.

Experimentally the strength of the Lane potential and its momentum
dependence can be extracted from the nucleon-nucleus scattering
data and/or (p,n) charge exchange reactions. Earlier optical model
analyses of the experimental data with incident energies between 7 and 100 MeV
indicate that the $(U_n-U_p)/2\beta$ at normal nuclear matter
density has a value of about $28\pm 6$ MeV at $k=0$ and decreases
as a function of incident energy with a slope between 0.1 to
0.2\cite{data,hofmann}.

The Lane potential is represented by the dashed area, as results
extracted from the experimental data of nucleon-nucleus scattering
based on the optical potential model~\cite{data}. This area is
crossed by both the BHF symmetry potential for $\beta=0.2$ and the
bulk contribution of the empirical one from Ref.~\cite{opt}
discussed later. The predicted strength of the isospin splitting
at $k=0$ is about $25$ MeV in good agreement with the empirical
value $22\sim 34$ MeV, extracted from the experimental data of
nucleon-nucleus scattering based on the optical potential
model~\cite{data,opt}.
\begin{figure}[tbh]

\caption{Comparison among different symmetry potentials: the
dashed area is the Lane potential $a-bE_{kin}$ with $a=22-34$ MeV
and $b=0.1-0.2$(see Ref.~\cite{bao} for more details), the dashed
line is the BHF result for asymmetric nuclear matter with
$\rho=0.16 fm^{-3}$ and $\beta=0.2$ and the solid curve is from
empirical parametrization of nucleon-$^{208}$Pb
scattering\cite{opt}.}
 \label{usym}
\end{figure}

A recent investigation~\cite{opt} on a broad range of mass
($24\leq A\leq 209$) and incident energy (1 KeV $\leq E\leq $ 200
MeV) provides a new parametrization of the optical model potential
in terms of volume, surface, spin-orbit and Coulomb contributions.
Therefore, it should be more suitable than the one of
Fig.~\ref{usym} for a comparison with the microscopic potential of
nuclear matter, including the density, isospin and energy
dependence. Since a direct comparison does not take into account
the density variation in the nuclear surface, we have folded the
nuclear-matter mean field with the density profile of different
nuclei obtained from a Thomas-Fermi approximation\cite{Ring}. So
doing we can compare the folded mean field with the optical
potential as a function of $\beta=(N-Z)/A$. The results are
plotted in Fig.~\ref{dunp} for incident energy equal to the Fermi
energy. The theoretical predictions slightly overestimate the
empirical ones, but the isospin shift turns out to be nicely
reproduced. Actually the Thomas-Fermi approximation works better
with heavy nuclei for which the comparison looks much better. The
momentum dependent optical-model potential of Ref.\cite{opt} gets
out of the dashed area at low energy, indicating that some
uncertainties still affect the optical-model parametrizations.

\begin{center}
\begin{figure}[tbh]
\caption{Isospin shift of empirical optical potential at the Fermi
energy (volume term only) for several nuclei(symbols) in
comparison with the theoretical nuclear-matter BHF predictions
(big triangles joined by dashed lines). Small symbols
are the Optical potential fit by Koning and Delaroche (NPA 2003). }%
\label{dunp}%
\end{figure}
\end{center}

\begin{center}
\begin{figure}[tbh]
\caption{Comparison of the BHF symmetry potential with
other predictions (see text).}%
\label{comp}%
\end{figure}
\end{center}

At high density and energy one may have to rely on the scarce
information from HIC induced by high energy radioactive beams. In
this regard, it is interesting to mention that isospin diffusion
has been found to be rather sensitive to the momentum-dependence
of $U_{\rm sym}$~\cite{chen04}. Up to now only the
phenomenological parametrizations of the momentum-dependent
symmetry potential has been adopted in the dynamical simulations
of heavy ion collisions, therefore it is instructive to make a
comparison between the present microscopic symmetry potential with
the phenomenological ones\cite{das:2003}. In Fig.~\ref{comp} is
plotted the symmetry potentials versus momentum for three values
of density $\rho=0.085, 0.17$ and 0.34 fm$^{-3}$. In the figure
the curves with filled symbols are the results from the BHF
calculations, repeated in the four panels, while the curves with
open symbols are the phenomenological ones of Ref.\cite{das:2003}
and different panels correspond to different parametrizations. It
is clear from the figure that the microscopic $U_{\rm sym}$ shows
a remarkably different behavior from the phenomenological ones as
a function of density and momentum. All of the four
phenomenological symmetry potentials drop much faster at high
density $\rho=0.34$ fm$^{-3}$ as the momentum increases as
compared to our BHF one. In the two cases of the GBD(0) and GBD(1)
parametrization, the deviation from the BHF prediction is
especially large in the whole density and momentum regions
considered here. At the normal nuclear density, the momentum
dependence of the Gogny and MDI(0) parametrization is closer to
our microscopic one, but the Gogny $U_{\rm sym}$ presents an
opposite density dependence and discrepant dramatically with the
microscopic one at high densities. For example, at $\rho=0.34$
fm$^{-3}$ the Gogny $U_{\rm sym}$ is attractive (repulsive) for
neutrons (protons), while the BHF one is repulsive (attractive)
for neutrons (protons) up to $k\simeq 4$ fm$^{-1}$. Even the
MDI(0) parametrization which is closest to our microscopic
prediction, turns out to become quite different at high densities.

As the last point we discuss the effect of the rearrangement
contribution of the TBF. Due to the density dependence of the
effective force ${\tilde v}$, the TBF provides an extra repulsive
contribution $\Sigma_{tbf}$ (see Eq.(\ref{eq:sigma})) to both the
proton and neutron s.p. potentials. At high density this
contribution is expected to be strongly momentum dependent and may
affect considerably the high momentum components of the
fragmentation residues in HIC. We compare our calculated results
with the parametrization of the optical potential used in the
transport-model simulations of elliptic flows in central
HIC~\cite{danielewicz:2000}, where high densities are reached. The
potentials in symmetric nuclear matter at $\rho=0.3$fm$^{-3}$ are
shown in Fig.\ref{hden}, where the line with full squares is the
one of Ref.~\cite{danielewicz:2000} which has been shown to
describe the observed elliptic flow data fairly well. It is clear
from the figure that the BHF potential without the TBF is too
attractive, especially at high densities, as compared to the one
proposed in Ref.~\cite{danielewicz:2000}, and its momentum
dependence at high momenta turns out to be too
weak~\cite{danielewicz:2000} for describing the experimental
elliptic flow data. Inclusion of the TBF effect only via the
$G$-matrix, i.e., inclusion of the TBF effect in the first two
terms of Eq.(\ref{eq:sigma}), weakens the dependence of the s.p
potential on momentum\cite{zuo:2002}. It is seen from
Fig.~\ref{hden} that the rearrangement contribution of the TBF,
i.e., the third term $\Sigma_{tbf}$ of Eq.(\ref{eq:sigma}),
provides a strongly extra repulsion to the optical potential and
improves remarkably the agreement between our microscopic
potential and the parametrized one of
Ref.~\cite{danielewicz:2000}. We find that the TBF rearrangement
leads to also a strong momentum dependence at high densities. For
instance at $\rho=0.3$ fm$^{-3}$ the extra term turns out to be
$\Delta U = 5.68 + 6.84 k^2$ in units of MeV.

To study the effect of the TBF rearrangement on the isospin
symmetry potential, we report in Fig.~\ref{dib} the results in
comparison with the recent DBHF predictions from
Refs.~\cite{dalen:2004,samma}. Therein it is displayed the neutron
s.p. potential (left-upper panel), the proton s.p. potential
(left-lower panel) and the symmetry potential (right panel) in
neutron-rich matter with $\beta=0.4$ for both cases with and
without including the $\Sigma_{tbf}$ term. It is seen that the BHF
neutron and proton s.p. potentials without the $\Sigma_{tbf}$ term
are much more attractive than the DBHF ones. Inclusion of the
$\Sigma_{tbf}$ term leads to a strong enhancement of the repulsion
of both the proton and neutron s.p. potentials and reduces
substantially the disagreement between the s.p. potentials
predicted by the nonrelativistic BHF and the relativistic DBHF
approaches. It is shown in the right panel of Fig.~\ref{dib} that
the effect of the $\Sigma_{tbf}$ term on the isospin symmetry
potential is very weak, indicting that the contributions of the
$\Sigma_{tbf}$ term to the neutron and proton potentials cancel
out in a wide momentum range with each other to a large extent. In
both cases with and without the $\Sigma_{tbf}$ term, the isospin
symmetry potentials obtained by the BHF approach display an
overall agreement with those by the DBHF approach. At high
momentum our symmetry potential is slight lower than the DBHF
ones. The difference between the two DBHF calculations may be
attributed to the different methods adopted to extract the
selfenergy as discussed in Ref.~\cite{sehn:1997} where it is shown
that the determination of the nucleon selfenergy in the DBHF
framework is still affected by some uncertainties.

\begin{center}
\begin{figure}[tbh]

 \caption{Optical potential in nuclear matter as a function of
nucleon energy at density $\rho=0.3 fm^{-3}$ from the BHF
calculation including all the three terms of Eq.\ref{eq:sigma}
with the Argonne $V_{18}$ interaction plus TBF (solid line). The
squares are from Ref.\cite{danielewicz:2000}. The circles are
from the earlier BHF calculation without three-body force~\cite{baldo:1989}.}%
\label{hden}%
\end{figure}
\end{center}

\begin{figure}[tbh]
\vglue 1.5cm
\caption{Comparison of the BHF symmetry potential with and without
the $\Sigma_{tbf}$ contribution with Dirac-Brueckner predictions.
The curve with full circles is
from Ref.~\cite{samma}, the one with full squares is from Ref.\cite{dalen:2004}}%
\label{dib}%
\end{figure}

\section{Summary and conclusions}
\label{summary}

In summary, we have investigated the momentum and density
dependence of the symmetry potential and discussed the origin of
the neutron-proton effective mass splitting in neutron-rich
nuclear matter within the framework of the Brueckner theory.

We have found that the isospin behavior of the momentum dependent
neutron and proton s.p. potentials can be traced back to the
effect of the $SD$ tensor component of the NN interaction and,
consequently, the neutron-proton effective mass splitting is
essentially determined by the intrinsic properties of the NN
interaction. The obtained neutron-proton effective mass splitting
in neutron-rich matter is $m^*_n>m^*_p$ in good agreement with the
recent predictions by the non relativistic limit of DBHF
approach~\cite{dalen:2005}. The isospin splitting $m^*_n>m^*_p$ is
shown to stem from the splitting of the $k$-mass, i.e., from the
spatial nonlocality of the microscopic neutron and proton s.p.
potentials.

It turns out that the predicted symmetry potential depends
sensitively on density and momentum, but almost independent of the
isospin asymmetry (Lane ansatz). In the energy and density regions
most relevant for the nucleus-nucleus scattering up to an incident
energy of about 300 MeV per nucleon, our microscopic symmetry
potential is repulsive for neutrons and attractive for protons,
and its strength becomes smaller as momentum increases for a fixed
density. A satisfactory support to the microscopic predictions is
provided by a recent optical model parametrizations of
nucleon-nucleus scattering~\cite{opt}. In dense nuclear matter,
which can be probed in HIC at intermediate and high energies, the
symmetry potential turns out to become stronger in the high
momentum region up to about 4 fm$^{-1}$ as increasing density. At
high energy relativistic effects manifest with a strong momentum
dependence~\cite{danielewicz:2000,dalen:2005}, which can be
interpreted as an effect of the TBF. At the present time a
comparison with phenomenological Skyrme-like or Gogny predictions
of the symmetry potential~\cite{bao} is quite difficult, and
empirical constraints on their parameters are needed to make it
useful.

In the BHF approach, the TBF contribution has been included by
reducing the TBF to an equivalent effective two-body force. The
rearrangement effect due to the density dependence of the
equivalent force has been found to provide an extra repulsive
contribution to the proton and neutron s.p. potentials, which
improves substantially the agreement of our nonrelativistic s.p.
potential with the parametrized potential for describing the
elliptic flow data\cite{danielewicz:2000} and those predicted by
the DBHF approach~\cite{dalen:2004,samma}. The TBF rearrangement
has been found to affect only slightly affect the isospin symmetry
potential due to the cancellation between the two contributions to
the proton and neutron potentials. Our calculated symmetry
potential is shown to be in an overall agreement with the DBHF
predictions.

Physical observables which are sensitive to the symmetry
potential, including the neutron-to-proton ratio of
pre-equilibrium nucleon emission, neutron-to-proton differential
flow and isospin diffusion, are expected to provide experimental
constraints on the momentum and density dependence of the symmetry
potential~\cite{bao}.

\section{Acknowledgment}

We thank Dr. P. Danielewicz for valuable comments. One of us (W.
Zuo) acknowledges the warm hospitality extended to him at LNS-INFN
(Catania) where this work has been carried out. The work was done
within the Asia-Link project{'\it Nuclear Physics and
Astrophysics'} (CN/ASIA-LINK/008(94791)) granted by the European
Commission. The work of W. Zuo was supported in part by the Major
Prophase Research Project of Fundamental Research of the Ministry
of Science and Technology of China (2002CCB00200), the Chinese
Major State Basic Research Development Program (G2000077400), the
Knowledge Innovative Project of CAS (KJCX2-SW-N02), and the
National Natural Science Foundation of China (10235030). The work
of Bao-An Li was supported in part by the US National Science
Foundation under grants PHY0243571, PHY0354572 and the
NASA-Arkansas Space Grants Consortium Award ASU15154.

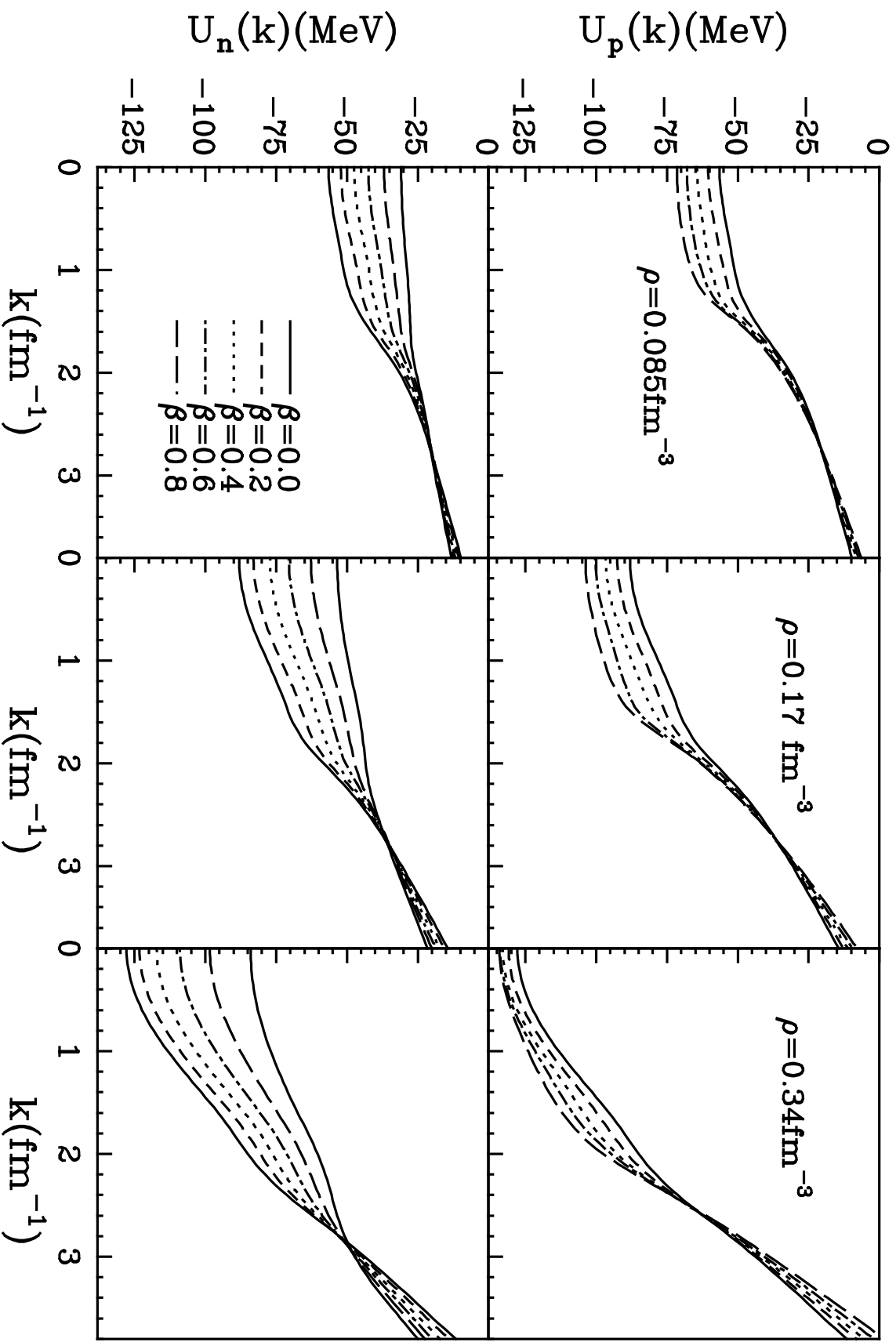
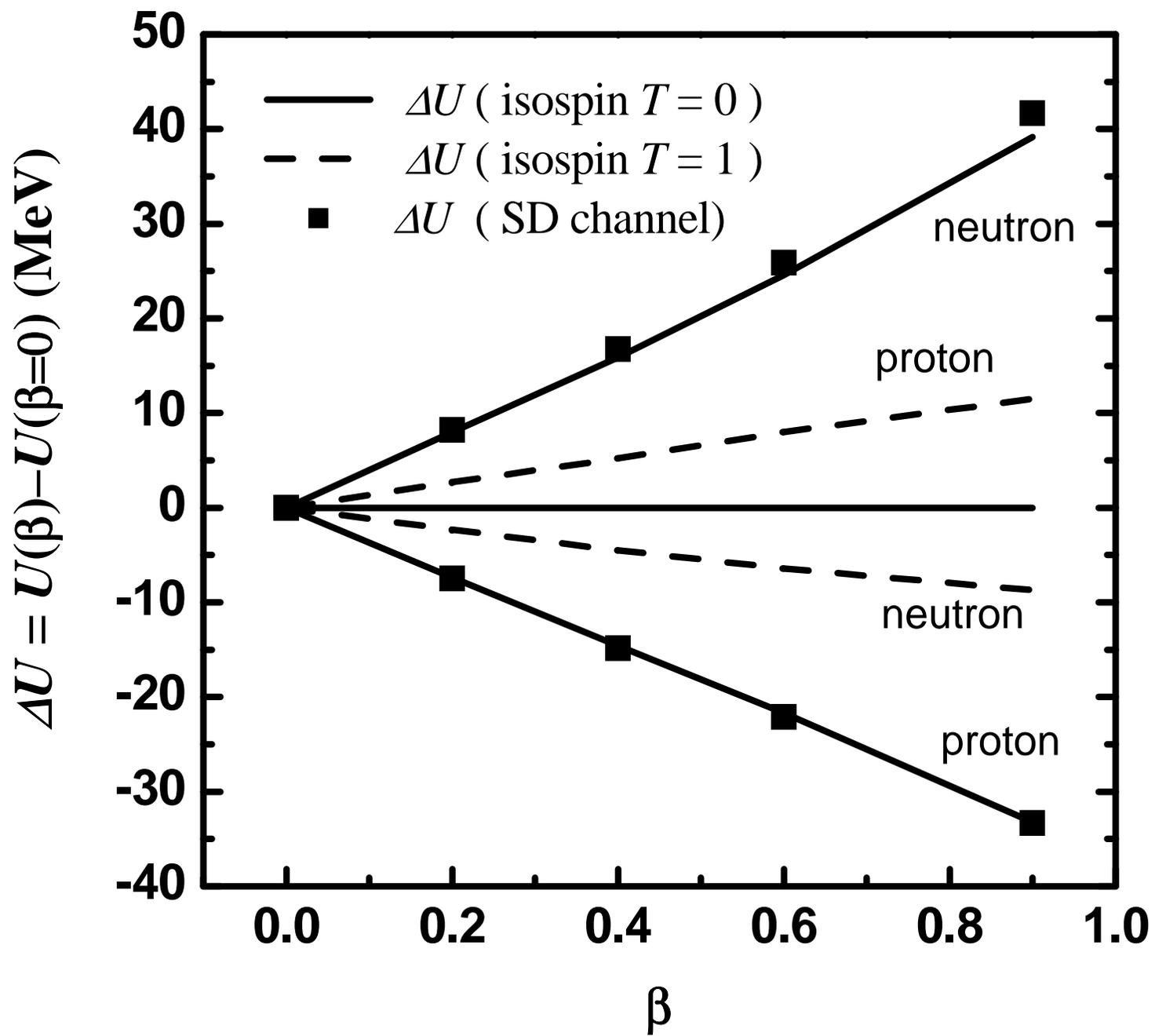
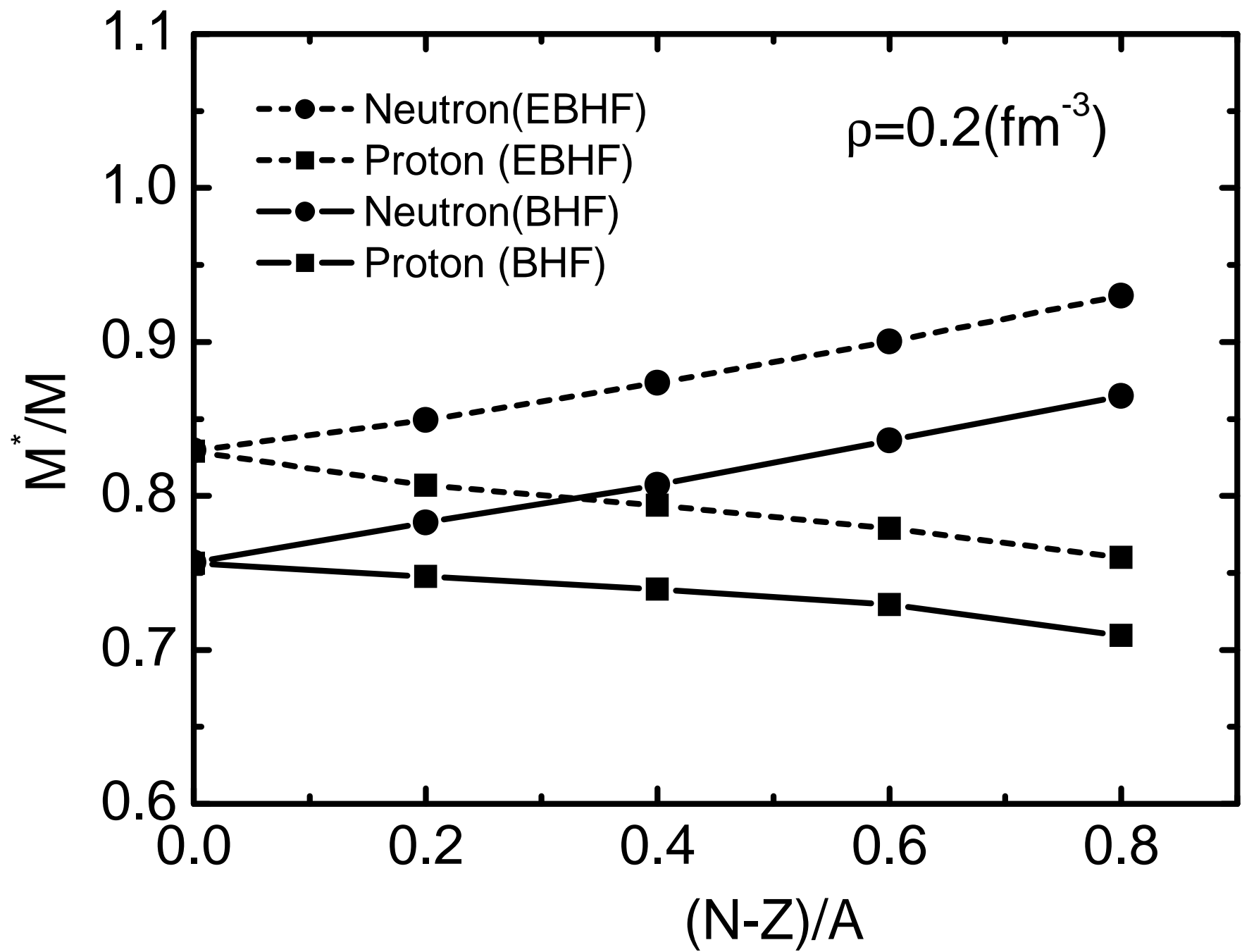
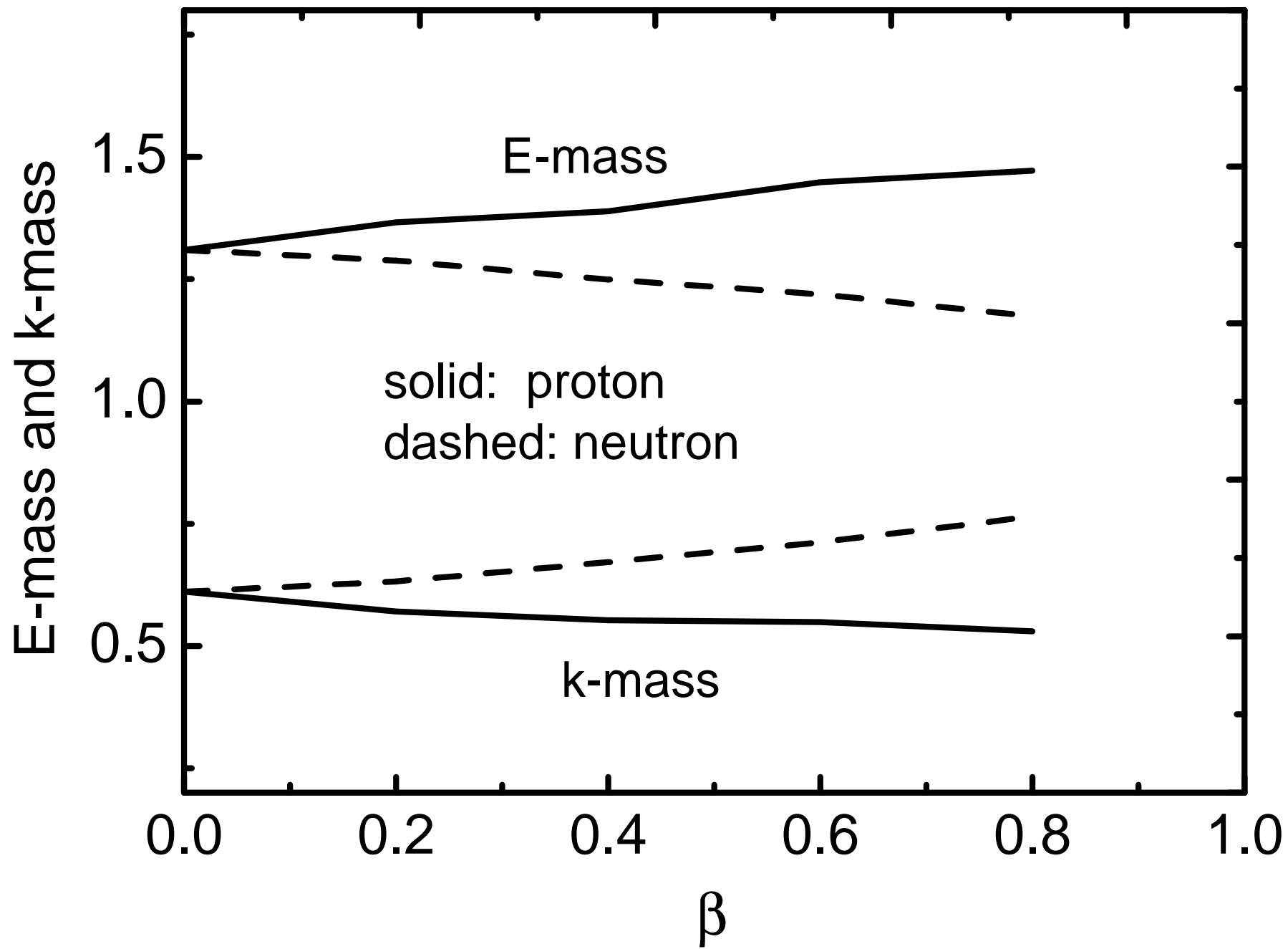
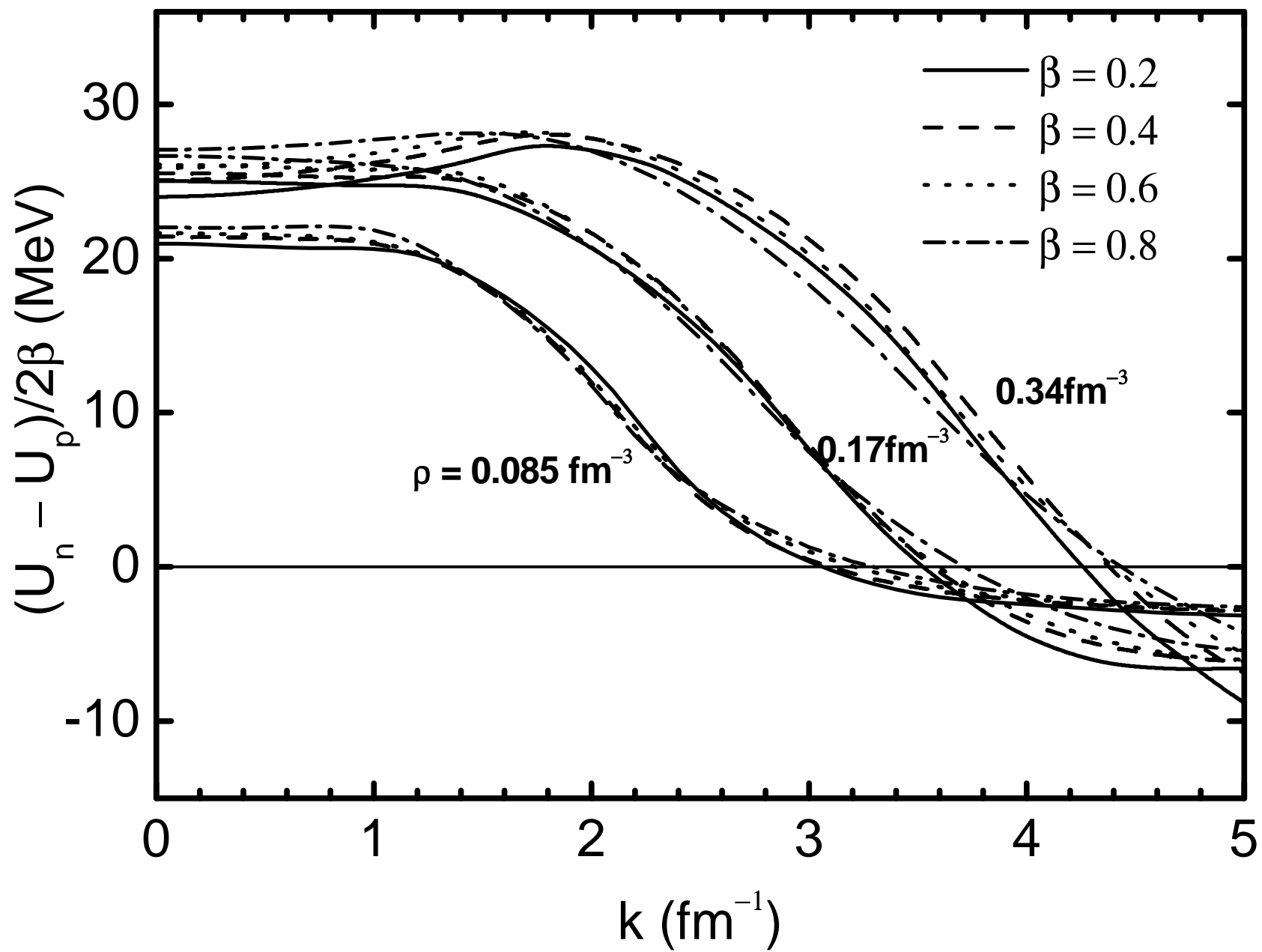
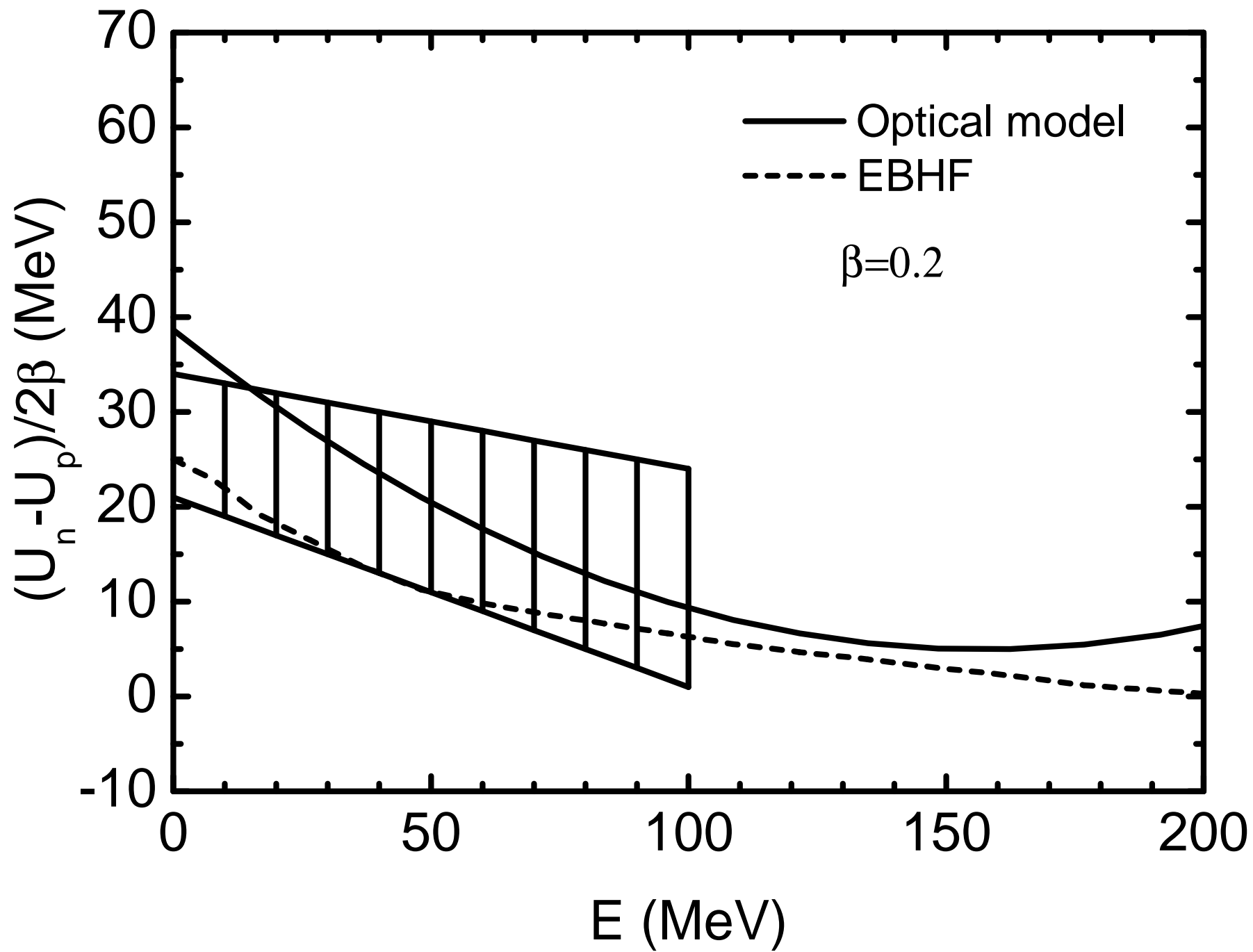
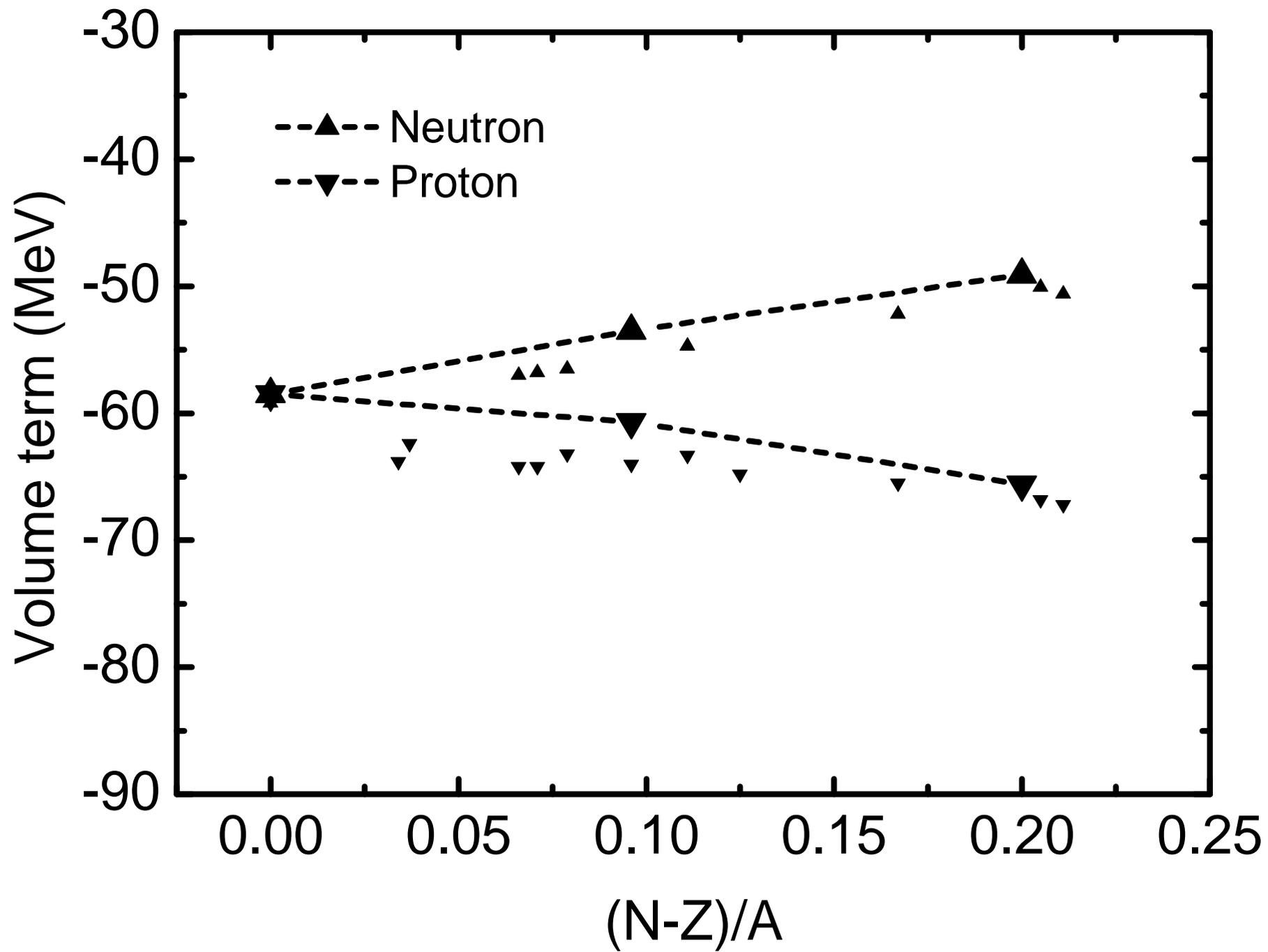
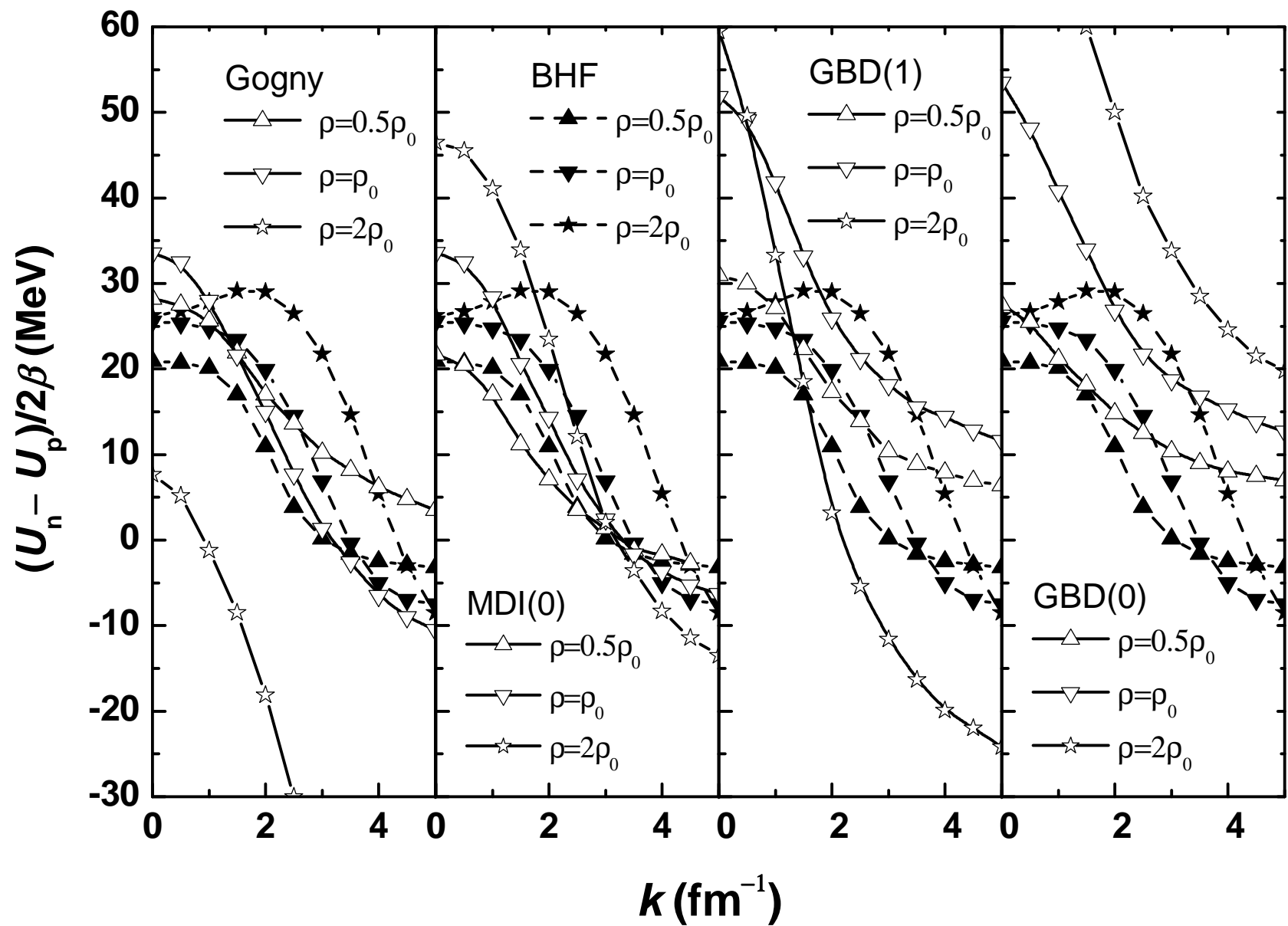
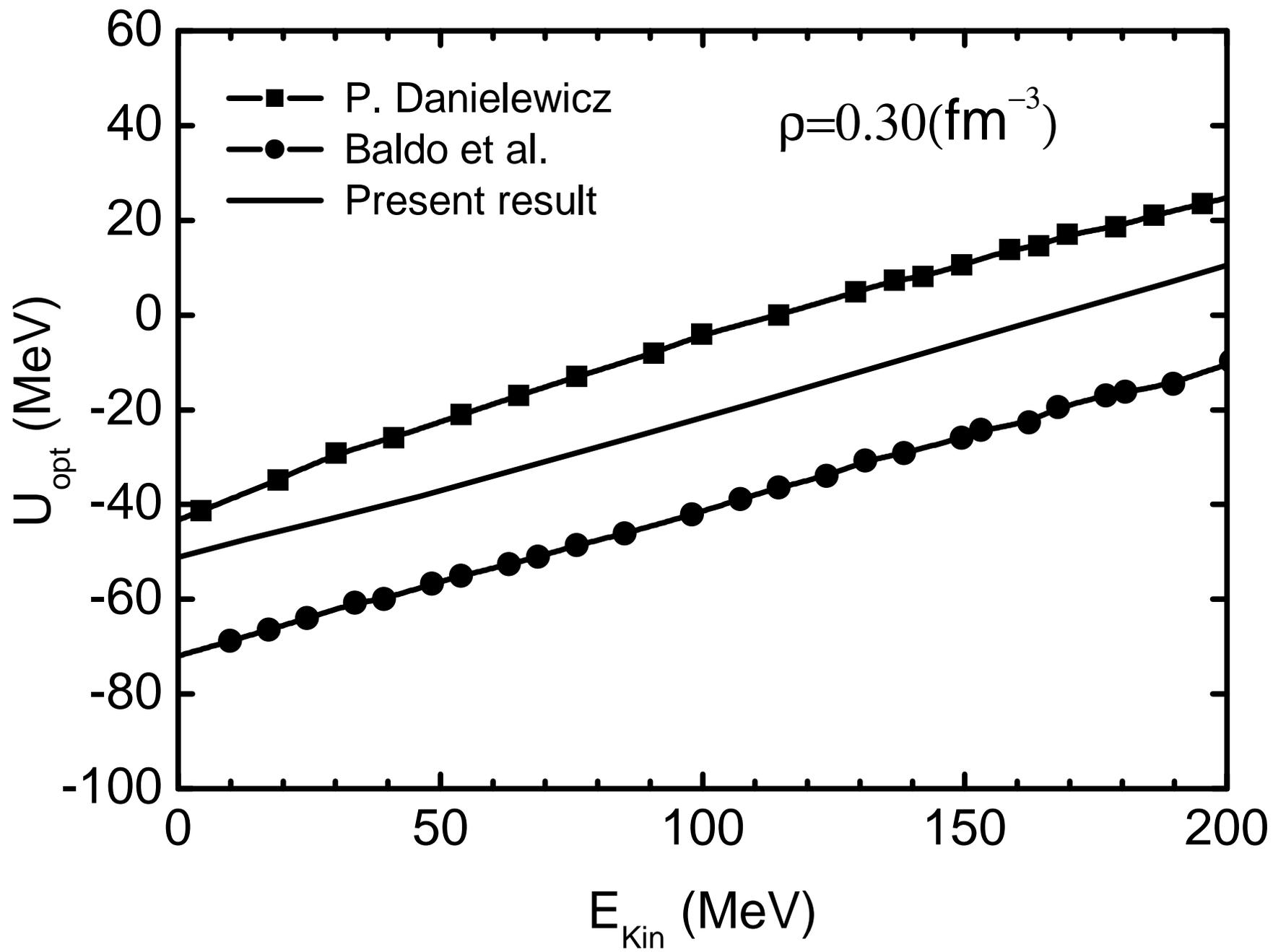
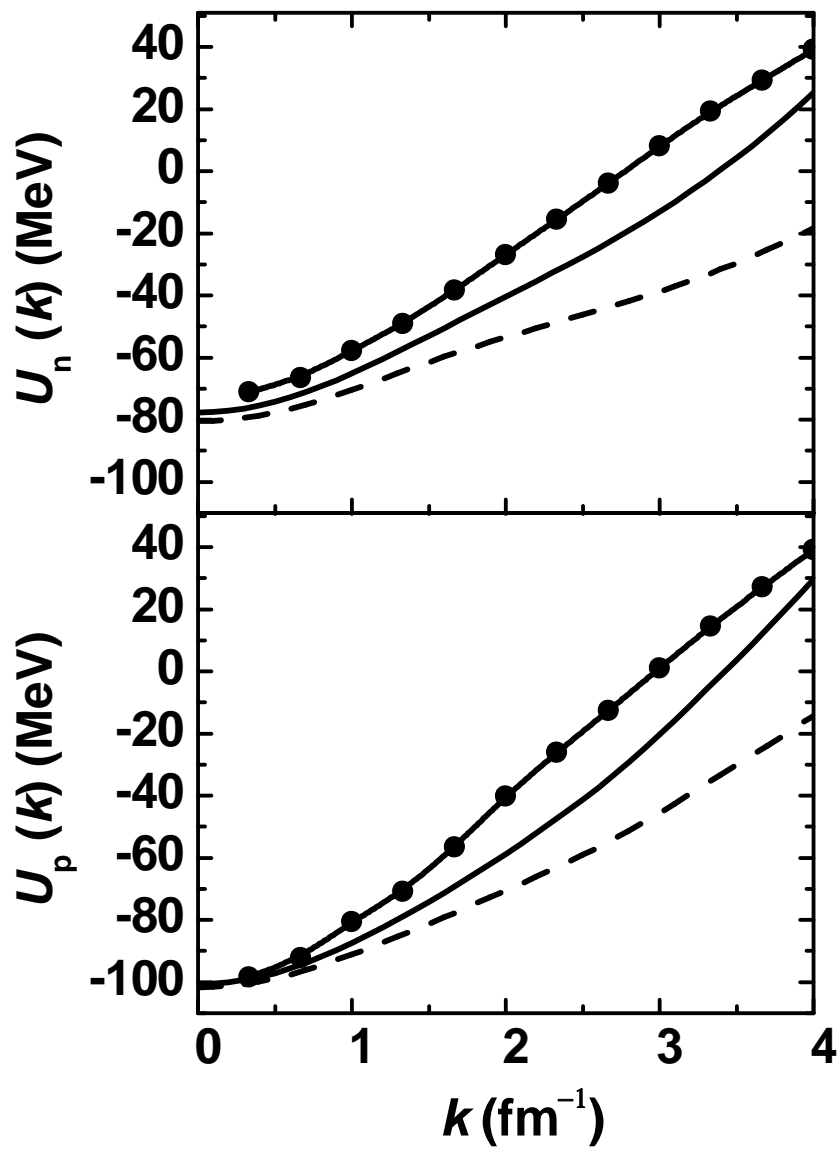
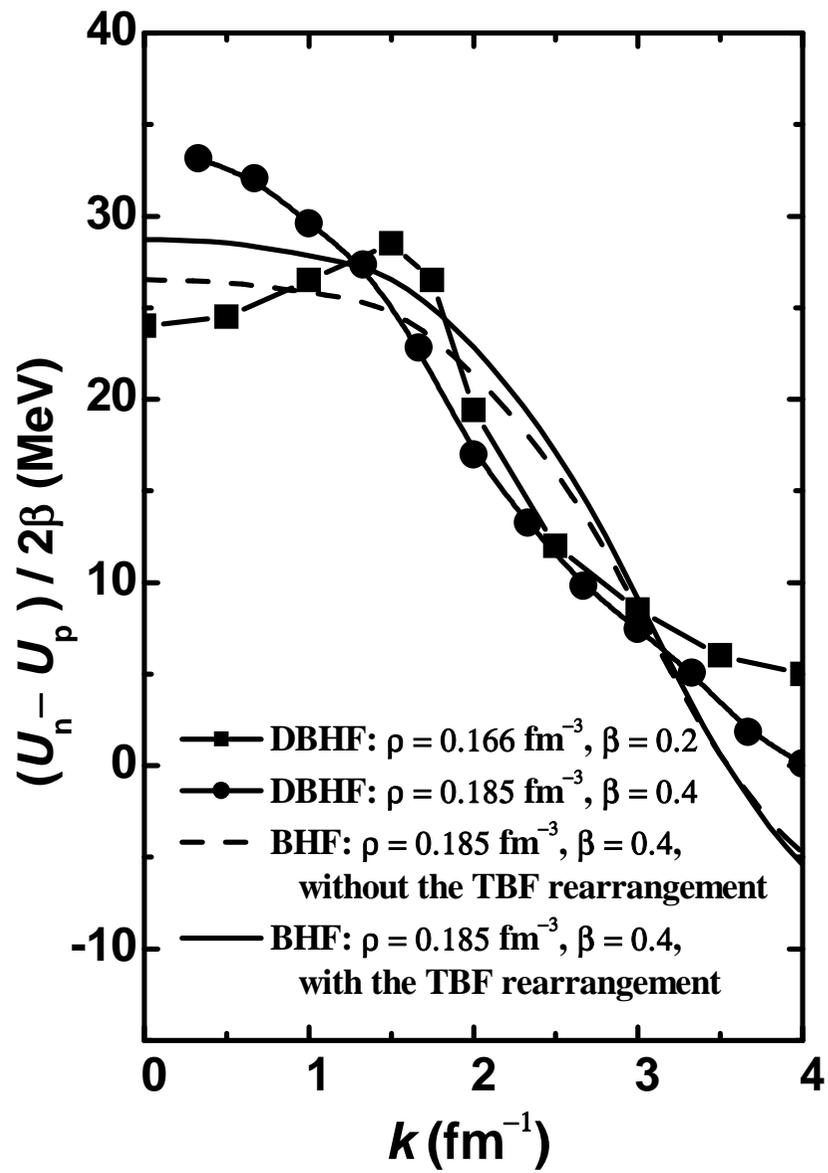
\end{document}